%%%%%%%%%%%%%%%%%%%%%%%%%%%%%%% including param.2 %%%%%%%%%%%%%%%%%%%%%%%%%%%%%%%
%param.2
\magnification=\magstep1
\def\firstpage{1}
\pageno=\firstpage
%%%%%%%%%%%%%%%%%%%%%%%%%%%%%%%%% end of param.2 %%%%%%%%%%%%%%%%%%%%%%%%%%%%%%%%
%%%%%%%%%%%%%%%%%%%%%%%%%%%%%%% including fonts.7 %%%%%%%%%%%%%%%%%%%%%%%%%%%%%%%
%fonts.6
\font\fiverm=cmr5
\font\sevenrm=cmr7

\font\eightrm=cmr8
\font\eightbf=cmbx8
\font\ninerm=cmr9
\font\ninebf=cmbx9
\font\tenbf=cmbx10
\font\twelvebf=cmbx12
%

%

%

%
%%%%%%%%%%%%%%%%%%%%%%%%%%%%%%%%% end of fonts.7 %%%%%%%%%%%%%%%%%%%%%%%%%%%%%%%%
%%%%%%%%%%%%%%%%%%%%%%%%%%%%%%% including smallfonts.tex %%%%%%%%%%%%%%%%%%%%%%%%%%%%%%%
%smallfonts.tex
%
\newskip\ttglue
\font\fiverm=cmr5
\font\fivei=cmmi5
\font\fivesy=cmsy5
\font\fivebf=cmbx5
\font\sixrm=cmr6
\font\sixi=cmmi6
\font\sixsy=cmsy6
\font\sixbf=cmbx6
\font\sevenrm=cmr7
\font\eightrm=cmr8
\font\eighti=cmmi8
\font\eightsy=cmsy8
\font\eightit=cmti8
\font\eightsl=cmsl8
\font\eighttt=cmtt8
\font\eightbf=cmbx8
\font\ninerm=cmr9
\font\ninei=cmmi9
\font\ninesy=cmsy9
\font\nineit=cmti9
\font\ninesl=cmsl9
\font\ninett=cmtt9
\font\ninebf=cmbx9
\font\twelverm=cmr12
\font\twelvei=cmmi12
\font\twelvesy=cmsy12
\font\twelveit=cmti12
\font\twelvesl=cmsl12
\font\twelvett=cmtt12
\font\twelvebf=cmbx12

%% EIGHT POINT FONT FAMILY

\def\eightpoint{\def\rm{\fam0\eightrm}
  \textfont0=\eightrm \scriptfont0=\sixrm \scriptscriptfont0=\fiverm
  \textfont1=\eighti  \scriptfont1=\sixi  \scriptscriptfont1=\fivei
  \textfont2=\eightsy  \scriptfont2=\sixsy  \scriptscriptfont2=\fivesy
  \textfont3=\tenex  \scriptfont3=\tenex  \scriptscriptfont3=\tenex
  \textfont\itfam=\eightit  \def\it{\fam\itfam\eightit}
  \textfont\slfam=\eightsl  \def\sl{\fam\slfam\eightsl}
  \textfont\ttfam=\eighttt  \def\tt{\fam\ttfam\eighttt}
  \textfont\bffam=\eightbf  \scriptfont\bffam=\sixbf
    \scriptscriptfont\bffam=\fivebf  \def\bf{\fam\bffam\eightbf}
  \tt  \ttglue=.5em plus.25em minus.15em
  \normalbaselineskip=9pt
  \setbox\strutbox=\hbox{\vrule height7pt depth2pt width0pt}
  \let\sc=\sixrm  \let\big=\eightbig \normalbaselines\rm}

\def\eightbig#1{{\hbox{$\textfont0=\ninerm\textfont2=\ninesy
        \left#1\vbox to6.5pt{}\right.$}}}

%% NINE POINT FONT FAMILY

\def\ninepoint{\def\rm{\fam0\ninerm}
  \textfont0=\ninerm \scriptfont0=\sixrm \scriptscriptfont0=\fiverm
  \textfont1=\ninei  \scriptfont1=\sixi  \scriptscriptfont1=\fivei
  \textfont2=\ninesy  \scriptfont2=\sixsy  \scriptscriptfont2=\fivesy
  \textfont3=\tenex  \scriptfont3=\tenex  \scriptscriptfont3=\tenex
  \textfont\itfam=\nineit  \def\it{\fam\itfam\nineit}
  \textfont\slfam=\ninesl  \def\sl{\fam\slfam\ninesl}
  \textfont\ttfam=\ninett  \def\tt{\fam\ttfam\ninett}
  \textfont\bffam=\ninebf  \scriptfont\bffam=\sixbf
    \scriptscriptfont\bffam=\fivebf  \def\bf{\fam\bffam\ninebf}
  \tt  \ttglue=.5em plus.25em minus.15em
  \normalbaselineskip=11pt
  \setbox\strutbox=\hbox{\vrule height8pt depth3pt width0pt}
  \let\sc=\sevenrm  \let\big=\ninebig \normalbaselines\rm}

\def\ninebig#1{{\hbox{$\textfont0=\tenrm\textfont2=\tensy
        \left#1\vbox to7.25pt{}\right.$}}}

%% TWELVE POINT FONT FAMILY --- not really small

\def\twelvepoint{\def\rm{\fam0\twelverm}
  \textfont0=\twelverm \scriptfont0=\eightrm \scriptscriptfont0=\sixrm
  \textfont1=\twelvei  \scriptfont1=\eighti  \scriptscriptfont1=\sixi
  \textfont2=\twelvesy  \scriptfont2=\eightsy  \scriptscriptfont2=\sixsy
  \textfont3=\tenex  \scriptfont3=\tenex  \scriptscriptfont3=\tenex
  \textfont\itfam=\twelveit  \def\it{\fam\itfam\twelveit}
  \textfont\slfam=\twelvesl  \def\sl{\fam\slfam\twelvesl}
  \textfont\ttfam=\twelvett  \def\tt{\fam\ttfam\twelvett}
  \textfont\bffam=\twelvebf  \scriptfont\bffam=\eightbf
    \scriptscriptfont\bffam=\sixbf  \def\bf{\fam\bffam\twelvebf}
  \tt  \ttglue=.5em plus.25em minus.15em
  \normalbaselineskip=11pt
  \setbox\strutbox=\hbox{\vrule height8pt depth3pt width0pt}
  \let\sc=\sevenrm  \let\big=\twelvebig \normalbaselines\rm}

\def\twelvebig#1{{\hbox{$\textfont0=\tenrm\textfont2=\tensy
        \left#1\vbox to7.25pt{}\right.$}}}
\catcode`\@=11
%
%  Include all definitions related to the fonts msam, msbm and eufm, so that
%  when this file is used by itself, the results with respect to those fonts
%  are equivalent to what they would have been using AMS-TeX.
%  Most symbols in fonts msam and msbm are defined using \newsymbol;
%  however, a few symbols that replace composites defined in plain must be
%  defined with \mathchardef.

\def\undefine#1{\let#1\undefined}
\def\newsymbol#1#2#3#4#5{\let\next@\relax
 \ifnum#2=\@ne\let\next@\msafam@\else
 \ifnum#2=\tw@\let\next@\msbfam@\fi\fi
 \mathchardef#1="#3\next@#4#5}
\def\mathhexbox@#1#2#3{\relax
 \ifmmode\mathpalette{}{\m@th\mathchar"#1#2#3}%
 \else\leavevmode\hbox{$\m@th\mathchar"#1#2#3$}\fi}
\def\hexnumber@#1{\ifcase#1 0\or 1\or 2\or 3\or 4\or 5\or 6\or 7\or 8\or
 9\or A\or B\or C\or D\or E\or F\fi}

\font\tenmsa=msam10
\font\sevenmsa=msam7
\font\fivemsa=msam5
\newfam\msafam
\textfont\msafam=\tenmsa
\scriptfont\msafam=\sevenmsa
\scriptscriptfont\msafam=\fivemsa
\edef\msafam@{\hexnumber@\msafam}
\mathchardef\dabar@"0\msafam@39
\def\dashrightarrow{\mathrel{\dabar@\dabar@\mathchar"0\msafam@4B}}
\def\dashleftarrow{\mathrel{\mathchar"0\msafam@4C\dabar@\dabar@}}

\def\ulcorner{\delimiter"4\msafam@70\msafam@70 }
\def\urcorner{\delimiter"5\msafam@71\msafam@71 }
\def\llcorner{\delimiter"4\msafam@78\msafam@78 }
\def\lrcorner{\delimiter"5\msafam@79\msafam@79 }
%    Note that there should not be a final space after the digits for a
%    \mathhexbox@.
\def\yen{{\mathhexbox@\msafam@55}}
\def\checkmark{{\mathhexbox@\msafam@58}}
\def\circledR{{\mathhexbox@\msafam@72}}
\def\maltese{{\mathhexbox@\msafam@7A}}

\font\tenmsb=msbm10
\font\sevenmsb=msbm7
\font\fivemsb=msbm5
\newfam\msbfam
\textfont\msbfam=\tenmsb
\scriptfont\msbfam=\sevenmsb
\scriptscriptfont\msbfam=\fivemsb
\edef\msbfam@{\hexnumber@\msbfam}
\def\Bbb#1{{\fam\msbfam\relax#1}}
\def\widehat#1{\setbox\z@\hbox{$\m@th#1$}%
 \ifdim\wd\z@>\tw@ em\mathaccent"0\msbfam@5B{#1}%
 \else\mathaccent"0362{#1}\fi}
\def\widetilde#1{\setbox\z@\hbox{$\m@th#1$}%
 \ifdim\wd\z@>\tw@ em\mathaccent"0\msbfam@5D{#1}%
 \else\mathaccent"0365{#1}\fi}
\font\teneufm=eufm10
\font\seveneufm=eufm7
\font\fiveeufm=eufm5
\newfam\eufmfam
\textfont\eufmfam=\teneufm
\scriptfont\eufmfam=\seveneufm
\scriptscriptfont\eufmfam=\fiveeufm

\catcode`\@=11
%%  Load amssym.def if necessary: If \newsymbol is undefined, do nothing
%%  and the following \input statement will be executed; otherwise
%%  change \input to a temporary no-op.
%#\ifx\undefined\newsymbol \else \begingroup\def\input#1 {\endgroup}\fi
%#\input amssym.def \relax
%%  Most symbols in fonts msam and msbm are defined using \newsymbol.  A few
%%  that are delimiters or otherwise require special treatment have already
%%  been defined as soon as the fonts were loaded.  Finally, a few symbols
%%  that replace composites defined in plain must be undefined first.
\newsymbol\boxdot 1200
\newsymbol\boxplus 1201
\newsymbol\boxtimes 1202
\newsymbol\square 1003
\newsymbol\blacksquare 1004
\newsymbol\centerdot 1205
\newsymbol\lozenge 1006
\newsymbol\blacklozenge 1007
\newsymbol\circlearrowright 1308
\newsymbol\circlearrowleft 1309
\undefine\rightleftharpoons
\newsymbol\rightleftharpoons 130A
\newsymbol\leftrightharpoons 130B
\newsymbol\boxminus 120C
\newsymbol\Vdash 130D
\newsymbol\Vvdash 130E
\newsymbol\vDash 130F
\newsymbol\twoheadrightarrow 1310
\newsymbol\twoheadleftarrow 1311
\newsymbol\leftleftarrows 1312
\newsymbol\rightrightarrows 1313
\newsymbol\upuparrows 1314
\newsymbol\downdownarrows 1315
\newsymbol\upharpoonright 1316
 
\newsymbol\downharpoonright 1317
\newsymbol\upharpoonleft 1318
\newsymbol\downharpoonleft 1319
\newsymbol\rightarrowtail 131A
\newsymbol\leftarrowtail 131B
\newsymbol\leftrightarrows 131C
\newsymbol\rightleftarrows 131D
\newsymbol\Lsh 131E
\newsymbol\Rsh 131F
\newsymbol\rightsquigarrow 1320
\newsymbol\leftrightsquigarrow 1321
\newsymbol\looparrowleft 1322
\newsymbol\looparrowright 1323
\newsymbol\circeq 1324
\newsymbol\succsim 1325
\newsymbol\gtrsim 1326
\newsymbol\gtrapprox 1327
\newsymbol\multimap 1328
\newsymbol\therefore 1329
\newsymbol\because 132A
\newsymbol\doteqdot 132B
 
\newsymbol\triangleq 132C
\newsymbol\precsim 132D
\newsymbol\lesssim 132E
\newsymbol\lessapprox 132F
\newsymbol\eqslantless 1330
\newsymbol\eqslantgtr 1331
\newsymbol\curlyeqprec 1332
\newsymbol\curlyeqsucc 1333
\newsymbol\preccurlyeq 1334
\newsymbol\leqq 1335
\newsymbol\leqslant 1336
\newsymbol\lessgtr 1337
\newsymbol\backprime 1038
\newsymbol\risingdotseq 133A
\newsymbol\fallingdotseq 133B
\newsymbol\succcurlyeq 133C
\newsymbol\geqq 133D
\newsymbol\geqslant 133E
\newsymbol\gtrless 133F
\newsymbol\sqsubset 1340
\newsymbol\sqsupset 1341
\newsymbol\vartriangleright 1342
\newsymbol\vartriangleleft 1343
\newsymbol\trianglerighteq 1344
\newsymbol\trianglelefteq 1345
\newsymbol\bigstar 1046
\newsymbol\between 1347
\newsymbol\blacktriangledown 1048
\newsymbol\blacktriangleright 1349
\newsymbol\blacktriangleleft 134A
\newsymbol\vartriangle 134D
\newsymbol\blacktriangle 104E
\newsymbol\triangledown 104F
\newsymbol\eqcirc 1350
\newsymbol\lesseqgtr 1351
\newsymbol\gtreqless 1352
\newsymbol\lesseqqgtr 1353
\newsymbol\gtreqqless 1354
\newsymbol\Rrightarrow 1356
\newsymbol\Lleftarrow 1357
\newsymbol\veebar 1259
\newsymbol\barwedge 125A
\newsymbol\doublebarwedge 125B
\undefine\angle
\newsymbol\angle 105C
\newsymbol\measuredangle 105D
\newsymbol\sphericalangle 105E
\newsymbol\varpropto 135F
\newsymbol\smallsmile 1360
\newsymbol\smallfrown 1361
\newsymbol\Subset 1362
\newsymbol\Supset 1363
\newsymbol\Cup 1264
 
\newsymbol\Cap 1265
 
\newsymbol\curlywedge 1266
\newsymbol\curlyvee 1267
\newsymbol\leftthreetimes 1268
\newsymbol\rightthreetimes 1269
\newsymbol\subseteqq 136A
\newsymbol\supseteqq 136B
\newsymbol\bumpeq 136C
\newsymbol\Bumpeq 136D
\newsymbol\lll 136E
 
\newsymbol\ggg 136F
 
\newsymbol\circledS 1073
\newsymbol\pitchfork 1374
\newsymbol\dotplus 1275
\newsymbol\backsim 1376
\newsymbol\backsimeq 1377
\newsymbol\complement 107B
\newsymbol\intercal 127C
\newsymbol\circledcirc 127D
\newsymbol\circledast 127E
\newsymbol\circleddash 127F
\newsymbol\lvertneqq 2300
\newsymbol\gvertneqq 2301
\newsymbol\nleq 2302
\newsymbol\ngeq 2303
\newsymbol\nless 2304
\newsymbol\ngtr 2305
\newsymbol\nprec 2306
\newsymbol\nsucc 2307
\newsymbol\lneqq 2308
\newsymbol\gneqq 2309
\newsymbol\nleqslant 230A
\newsymbol\ngeqslant 230B
\newsymbol\lneq 230C
\newsymbol\gneq 230D
\newsymbol\npreceq 230E
\newsymbol\nsucceq 230F
\newsymbol\precnsim 2310
\newsymbol\succnsim 2311
\newsymbol\lnsim 2312
\newsymbol\gnsim 2313
\newsymbol\nleqq 2314
\newsymbol\ngeqq 2315
\newsymbol\precneqq 2316
\newsymbol\succneqq 2317
\newsymbol\precnapprox 2318
\newsymbol\succnapprox 2319
\newsymbol\lnapprox 231A
\newsymbol\gnapprox 231B
\newsymbol\nsim 231C
\newsymbol\ncong 231D
\newsymbol\diagup 201E
\newsymbol\diagdown 201F
\newsymbol\varsubsetneq 2320
\newsymbol\varsupsetneq 2321
\newsymbol\nsubseteqq 2322
\newsymbol\nsupseteqq 2323
\newsymbol\subsetneqq 2324
\newsymbol\supsetneqq 2325
\newsymbol\varsubsetneqq 2326
\newsymbol\varsupsetneqq 2327
\newsymbol\subsetneq 2328
\newsymbol\supsetneq 2329
\newsymbol\nsubseteq 232A
\newsymbol\nsupseteq 232B
\newsymbol\nparallel 232C
\newsymbol\nmid 232D
\newsymbol\nshortmid 232E
\newsymbol\nshortparallel 232F
\newsymbol\nvdash 2330
\newsymbol\nVdash 2331
\newsymbol\nvDash 2332
\newsymbol\nVDash 2333
\newsymbol\ntrianglerighteq 2334
\newsymbol\ntrianglelefteq 2335
\newsymbol\ntriangleleft 2336
\newsymbol\ntriangleright 2337
\newsymbol\nleftarrow 2338
\newsymbol\nrightarrow 2339
\newsymbol\nLeftarrow 233A
\newsymbol\nRightarrow 233B
\newsymbol\nLeftrightarrow 233C
\newsymbol\nleftrightarrow 233D
\newsymbol\divideontimes 223E
\newsymbol\varnothing 203F
\newsymbol\nexists 2040
\newsymbol\Finv 2060
\newsymbol\Game 2061
\newsymbol\mho 2066
\newsymbol\eth 2067
\newsymbol\eqsim 2368
\newsymbol\beth 2069
\newsymbol\gimel 206A
\newsymbol\daleth 206B
\newsymbol\lessdot 236C
\newsymbol\gtrdot 236D
\newsymbol\ltimes 226E
\newsymbol\rtimes 226F
\newsymbol\shortmid 2370
\newsymbol\shortparallel 2371
\newsymbol\smallsetminus 2272
\newsymbol\thicksim 2373
\newsymbol\thickapprox 2374
\newsymbol\approxeq 2375
\newsymbol\succapprox 2376
\newsymbol\precapprox 2377
\newsymbol\curvearrowleft 2378
\newsymbol\curvearrowright 2379
\newsymbol\digamma 207A
\newsymbol\varkappa 207B
\newsymbol\Bbbk 207C
\newsymbol\hslash 207D
\undefine\hbar
\newsymbol\hbar 207E
\newsymbol\backepsilon 237F
%  Restore the catcode value for @ that was previously saved.
%#\catcode`\@=\csname pre amssym.tex at\endcsname

%\endinput
%%%%%%%%%%%%%%%%%%%%%%%%%%%%%%%%% end of symbols.1 %%%%%%%%%%%%%%%%%%%%%%%%%%%%%%%%
%%%%%%%%%%%%%%%%%%%%%%%%%%%%%%% including links.1 %%%%%%%%%%%%%%%%%%%%%%%%%%%%%%%
% links.1
% adapted from http://insti.physics.sunysb.edu/~siegel/tex.shtml
%
% postscript/pdf
\newcount\marknumber	\marknumber=1
\newcount\countdp \newcount\countwd \newcount\countht
%
% for ordinary tex
%
\ifx\pdfoutput\undefined
\def\rgboo#1{}
\def\postscript#1{\special{" #1}}		% for dvips
\postscript{
	/bd {bind def} bind def
	/fsd {findfont exch scalefont def} bd
	/sms {setfont moveto show} bd
	/ms {moveto show} bd
	/pdfmark where		% printers ignore pdfmarks
	{pop} {userdict /pdfmark /cleartomark load put} ifelse
	[ /PageMode /UseOutlines		% bookmark window open
	/DOCVIEW pdfmark}
\def\bookmark#1#2{\postscript{		% #1=subheadings (if not 0)
	[ /Dest /MyDest\the\marknumber /View [ /XYZ null null null ] /DEST pdfmark
	[ /Title (#2) /Count #1 /Dest /MyDest\the\marknumber /OUT pdfmark}%
	\advance\marknumber by1}
\def\pdfclink#1#2#3{%
	\hskip-.25em\setbox0=\hbox{#2}%
		\countdp=\dp0 \countwd=\wd0 \countht=\ht0%
		\divide\countdp by65536 \divide\countwd by65536%
			\divide\countht by65536%
		\advance\countdp by1 \advance\countwd by1%
			\advance\countht by1%
		\def\linkdp{\the\countdp} \def\linkwd{\the\countwd}%
			\def\linkht{\the\countht}%
	\postscript{
		[ /Rect [ -1.5 -\linkdp.0 0\linkwd.0 0\linkht.5 ]
		/Border [ 0 0 0 ]
		/Action << /Subtype /URI /URI (#3) >>
		/Subtype /Link
		/ANN pdfmark}{\rgb{#1}{#2}}}
%
% for pdftex
%
\else
\def\rgboo#1{\pdfliteral{#1 rg #1 RG}}
\pdfcatalog{/PageMode /UseOutlines}		% bookmark window open
\def\bookmark#1#2{
	\pdfdest num \marknumber xyz
	\pdfoutline goto num \marknumber count #1 {#2}
	\advance\marknumber by1}
\def\pdfklink#1#2{%
	\noindent\pdfstartlink user
		{/Subtype /Link
		/Border [ 0 0 0 ]
		/A << /S /URI /URI (#2) >>}{\rgb{1 0 0}{#1}}%
	\pdfendlink}
\fi

\def\rgbo#1#2{\rgboo{#1}#2\rgboo{0 0 0}}
\def\rgb#1#2{\mark{#1}\rgbo{#1}{#2}\mark{0 0 0}}
\def\pdfklink#1#2{\pdfclink{1 0 0}{#1}{#2}}
\def\pdflink#1{\pdfklink{#1}{#1}}
%
% examples:
% \bookmark{0}{look here}
% \pdfclink{0 0 1}{testlink}{http://www.google.com/}
% \pdfklink{testlink}{http://www.google.com/}
% \pdflink{http://www.google.com/}
%%%%%%%%%%%%%%%%%%%%%%%%%%%%%%%%% end of links.1 %%%%%%%%%%%%%%%%%%%%%%%%%%%%%%%%
%%%%%%%%%%%%%%%%%%%%%%%%%%%%%%% including titles.9s %%%%%%%%%%%%%%%%%%%%%%%%%%%%%%%
%titles.8
% requires fonts.5 or higher and smallfonts.tex
% uses links.* if included
% enumerates \demo consecutively (no section number)
%
\newcount\seccount  %% sections
\newcount\subcount  %% subsection
\newcount\clmcount  %% claim
\newcount\equcount  %% equation
\newcount\refcount  %% reference
\newcount\demcount  %% example
\newcount\execount  %% exercise
\newcount\procount  %% problem
\seccount=0
\equcount=1
\clmcount=1
\subcount=1
\refcount=1
\demcount=0
\execount=0
\procount=0
%
%% MISC STUFF

\def\proofof(#1){\medskip\noindent{\bf Proof of \csname c#1\endcsname.\ }}

\def\references{\bigskip\noindent\hbox{\bf References}\medskip
                \ifx\pdflink\undefined\else\bookmark{0}{References}\fi}
\def\addref#1{\expandafter\xdef\csname r#1\endcsname{\number\refcount}
    \global\advance\refcount by 1}

\def\nextremark #1\par{\item{$\circ$} #1}
\def\firstremark #1\par{\bigskip\noindent{\bf Remarks.}
     \smallskip\nextremark #1\par}
\def\abstract#1\par{{\baselineskip=10pt
    \eightpoint\narrower\noindent{\eightbf Abstract.} #1\par}}
%
%% EQUATION
\def\equtag#1{\expandafter\xdef\csname e#1\endcsname{(\number\seccount.\number\equcount)}
              \global\advance\equcount by 1}
\def\equation(#1){\equtag{#1}\eqno\csname e#1\endcsname}
\def\equ(#1){\hskip-0.03em\csname e#1\endcsname}
%
%% CLAIMS (theorems etc)
\def\clmtag#1#2{\expandafter\xdef\csname cn#2\endcsname{\number\seccount.\number\clmcount}
                \expandafter\xdef\csname c#2\endcsname{#1~\number\seccount.\number\clmcount}
                \global\advance\clmcount by 1}
\def\claim #1(#2) #3\par{\clmtag{#1}{#2}
    \vskip.1in\medbreak\noindent
    {\bf \csname c#2\endcsname .\ }{\sl #3}\par
    \ifdim\lastskip<\medskipamount
    \removelastskip\penalty55\medskip\fi}
\def\clm(#1){\csname c#1\endcsname}
\def\clmno(#1){\csname cn#1\endcsname}
%
%% SECTION
\def\sectag#1{\global\advance\seccount by 1
              \expandafter\xdef\csname sectionname\endcsname{\number\seccount. #1}
              \equcount=1 \clmcount=1 \subcount=1 \execount=0 \procount=0}
\def\section#1\par{\vskip0pt plus.1\vsize\penalty-40
    \vskip0pt plus -.1\vsize\bigskip\bigskip
    \sectag{#1}
    \message{\sectionname}\leftline{\twelvebf\sectionname} %% was \magtenbf
    \nobreak\smallskip\noindent
    \ifx\pdflink\undefined
    \else
      \bookmark{0}{\sectionname}
    \fi}
%
%% SUBSECTION
\def\subtag#1{\expandafter\xdef\csname subsectionname\endcsname{\number\seccount.\number\subcount. #1}
              \global\advance\subcount by 1}
\def\subsection#1\par{\vskip0pt plus.05\vsize\penalty-20
    \vskip0pt plus -.05\vsize\medskip\medskip
    \subtag{#1}
    \message{\subsectionname}\leftline{\tenbf\subsectionname}
    \nobreak\smallskip\noindent
    \ifx\pdflink\undefined
    \else
      \bookmark{0}{.... \subsectionname}  %% can get a bit cluttered
    \fi}
%
%% DEMO (examples etc)
\def\demtag#1#2{\global\advance\demcount by 1
              \expandafter\xdef\csname de#2\endcsname{#1~\number\demcount}}
\def\demo #1(#2) #3\par{
  \demtag{#1}{#2}
  \vskip.1in\medbreak\noindent
  {\bf #1 \number\demcount.\enspace}
  {\rm #3}\par
  \ifdim\lastskip<\medskipamount
  \removelastskip\penalty55\medskip\fi}
\def\dem(#1){\csname de#1\endcsname}
%
%% EXERCISE
\def\exetag#1{\global\advance\execount by 1
              \expandafter\xdef\csname ex#1\endcsname{Exercise~\number\seccount.\number\execount}}
\def\exercise(#1) #2\par{
  \exetag{#1}
  \vskip.1in\medbreak\noindent
  {\bf Exercise \number\execount.}
  {\rm #2}\par
  \ifdim\lastskip<\medskipamount
  \removelastskip\penalty55\medskip\fi}
\def\exe(#1){\csname ex#1\endcsname}
%
%% PROBLEM
\def\protag#1{\global\advance\procount by 1
              \expandafter\xdef\csname pr#1\endcsname{\number\seccount.\number\procount}}
\def\problem(#1) #2\par{
  \ifnum\procount=0
    \parskip=6pt
    \vbox{\bigskip\centerline{\bf Problems \number\seccount}\nobreak\medskip}
  \fi
  \protag{#1}
  \item{\number\procount.} #2}
\def\pro(#1){Problem \csname pr#1\endcsname}
%
%%%%%%%%%%%%%%%%%%%%%%%%%%%%%%%%% end of titles.9s %%%%%%%%%%%%%%%%%%%%%%%%%%%%%%%%
%%%%%%%%%%%%%%%%%%%%%%%%%%%%%%% including macros.21 %%%%%%%%%%%%%%%%%%%%%%%%%%%%%%%
%macros.21
%
% requires fonts.5 or later
% also defines mathds (double strike) family
%
\def\rightheadline{\hfil}
\def\leftheadline{\sevenrm\hfil HANS KOCH\hfil}
\headline={\ifnum\pageno=\firstpage\hfil\else
\ifodd\pageno{{\fiverm\rightheadline}\number\pageno}
\else{\number\pageno\fiverm\leftheadline}\fi\fi}
\footline={\ifnum\pageno=\firstpage\hss\tenrm\folio\hss\else\hss\fi}

\let\cl=\centerline

\let\sss=\scriptscriptstyle

\def\GG{{\cal G}}
\def\HH{{\cal H}}

\def\LL{{\cal L}}

\def\PP{{\cal P}}
\def\QQ{{\cal Q}}

\def\SS{{\cal S}}

\def\UU{{\cal U}}

\def\ssE{{\sss E}}

\def\id{{\rm I}}

%
%%%%%%%%%%%%%%
\newfam\dsfam
\def\mathds #1{{\fam\dsfam\tends #1}}

\font\tends=dsrom10
\font\eightds=dsrom8
\textfont\dsfam=\tends
\scriptfont\dsfam=\eightds
%%%%%%%%%%%%%%
%

\def\integer{{\mathds Z}}

\def\real{{\mathds R}}
\def\complex{{\mathds C}}

\def\torus{{\Bbb T}}

\def\bcomma{\hbox{\bf ,}}

\def\twovec#1#2{\left[\matrix{#1\cr#2\cr}\right]}

\def\twomat#1#2#3#4{\left[\matrix{#1&#2\cr#3&#4\cr}\right]}

%

%

%

%
% from TeX book: used for commutative diagram
% in math mode, before using matrix, do
% \def\normalbaselines{\baselineskip20pt\lineskip3pt\lineskiplimit3pt}

%%%%%%%%%%%%%%%%%%%%%%%%%%%%%%%%% end of macros.21 %%%%%%%%%%%%%%%%%%%%%%%%%%%%%%%%
%%%%%%%%%%%%%%%%%%%%%%%%%%%%%%% including mygraphicx.tex %%%%%%%%%%%%%%%%%%%%%%%%%%%%%%%
%% modification of graphicx.tex by Nathan Goldschmidt
\input miniltx

\ifx\pdfoutput\undefined
  \def\Gin@driver{dvips.def}  % we are not running PDFTeX
\else
  \def\Gin@driver{pdftex.def} % we are running PDFTeX
\fi

\input graphicx.sty
\resetatcatcode
%%%%%%%%%%%%%%%%%%%%%%%%%%%%%%%%% end of mygraphicx.tex %%%%%%%%%%%%%%%%%%%%%%%%%%%%%%%%
%\input param.2
%\input fonts.7
%\input smallfonts.tex
%\input symbols.1
%\input links.1
%\input titles.9s
%\input macros.21
%\input mygraphicx.tex
%
\newdimen\savedparindent
\savedparindent=\parindent
\font\tenamsb=msbm10 \font\sevenamsb=msbm7 \font\fiveamsb=msbm5
\newfam\bbfam
\textfont\bbfam=\tenamsb
\scriptfont\bbfam=\sevenamsb
\scriptscriptfont\bbfam=\fiveamsb
\def\AM{{\ninerm AM~}}
\def\sAM{self-dual \AM}

\font\eighteufm=eufm8
\def\sbuR{{\hbox{\eighteufm R}}}
\def\buR{{\hbox{\teneufm R}}}
\def\circle{{\Bbb S}}
\def\mod{\mathop{\rm mod}\nolimits}
\def\mat{\mathop{\rm mat}\nolimits}
\def\rmSL{{\rm SL}}

\def\hdots{\line{\leaders\hbox to 0.5em{\hss .\hss}\hfil}}

\def\sfrac#1#2{\hbox{\raise2.2pt\hbox{$\scriptstyle#1$}\hskip-1.2pt
   {$\scriptstyle/$}\hskip-0.9pt\lower2.2pt\hbox{$\scriptstyle#2$}\hskip1.0pt}}
\def\shalf{\sfrac{1}{2}}

\def\stwovec#1#2{{\eightpoint\left[\matrix{#1\cr#2\cr}\right]}}
\def\stwomat#1#2#3#4{{\eightpoint\left[\matrix{#1&#2\cr#3&#4\cr}\right]}}
\def\today{\ifcase\month\or
January\or February\or March\or April\or May\or June\or
July\or August\or September\or October\or November\or December\fi
\space\number\day, \number\year}
\addref{Harp}
\addref{Hof}
\addref{AuAn}
\addref{BeSi}
\addref{TKNdN}
\addref{JoMo}
\addref{AvSi}
\addref{DyFa}
\addref{Lasti}
\addref{Lastii}
\addref{FaKa}
\addref{RuPi}
\addref{ATW}
\addref{MeOsWi}
\addref{OsAv}
\addref{CEY}
\addref{DaMe}
\addref{Puig}
\addref{GoSch}
\addref{AvJit}
\addref{AvG}
\addref{Satij}
\addref{KochAM}
\addref{KK}
\def\leftheadline{\sixrm\hfil Hans Koch and Sa\v sa Koci\'c\hfil\today}
\def\rightheadline{\sevenrm\hfil Renormalization and Hofstadter spectrum\hfil}
%
%%%%%%%%%%%%%%%%%%%%%%%%%%%%%%%%%%%%%%%%%%%%%%%%%%%%%%%%%%%%%%%%
\cl{{\twelvebf Renormalization and universality of the Hofstadter spectrum}}
\bigskip

\cl{
Hans Koch
\footnote{$^1$}
{\eightpoint\hskip-2.7em
Dept.~of Mathematics, The University of Texas at Austin,
Austin, TX 78712.}
and Sa\v sa Koci\'c
\footnote{$^2$}
{\eightpoint\hskip-2.7em
Dept.~of Mathematics,
The University of Mississippi, P.O.~Box 1848, University, MS 38677-1848.}
}

\bigskip
\abstract
We consider a renormalization transformation $\sbuR$
for skew-product maps of the type that
arise in a spectral analysis of the Hofstadter Hamiltonian.
Periodic orbits of $\sbuR$ determine universal constants
analogous to the critical exponents in the theory of phase transitions.
Restricting to skew-product maps over
a circle-rotations by the golden mean,
we find several periodic orbits for $\sbuR$,
and we conjecture that there are infinitely many.
Interestingly, all scaling factors that have been determined
to high accuracy appear to be algebraically related
to the circle-rotation number.
We present evidence that these values describe (among other things)
local scaling properties of the Hofstadter spectrum.

\section The Hofstadter model
%%%%%%%%%%%%%%%%%%%%%%%%%%%%%

The spectrum of the Hofstadter Hamiltonian [\rHarp,\rHof]
exhibits local self-similarity and scaling properties.
Using renormalization,
we argue that the scaling constant are universal,
and that many can be computed exactly.
Some of our results are rigorous,
while others are based on numerical computations
and hypotheses that remain to be verified.

The Hofstadter Hamiltonian describes Bloch electrons
moving on $\integer^2$
under the influence of a magnetic flux $2\pi\alpha$ through each unit cell.
It is given by
$$
H^\alpha=\lambda'(U_\alpha+U_\alpha^\ast)+\lambda(V_\alpha+V_\alpha^\ast)\,,
\qquad U_\alpha V_\alpha U_\alpha ^{-1}V_\alpha ^{-1}=e^{-2\pi i\alpha}\,,
\equation(HofstadterH)
$$
where $U_\alpha , V_\alpha$ are magnetic translations
and $\lambda , \lambda'$ are positive constants.
In the Landau gauge,
$(U_\alpha\phi)(n,m)=\phi(n-1,m)$
and $(V_\alpha\phi)(n,m)=e^{2\pi in\alpha}\phi(n,m-1)$.

For rational $\alpha=m/n$ with $m$ and $n$ coprime,
the spectrum of $H^\alpha$ consists of $n$ bands (closed intervals),
separated by gaps (open intervals), except at energy zero.
The width of these bands tend to zero as $n\to\infty$.
Numerically, the gaps are found to stay open
as $m/n\to\alpha$.
Another important observation [\rTKNdN] is the following.
Consider the integrated density of states
$d(\alpha,E)=\langle\delta_0,P^\alpha_{\ssE}\delta_0\rangle$,
where $P^\alpha_\ssE$ is the spectral projection for $H^\alpha$
associated with the interval $(-\infty,E]$,
and where $\delta_0$ is the Kronecker delta at the origin.
For each spectral gap
there exists an index $k\in\integer$, also known as the Hall conductance,
such that
$$
d(\alpha,E)\equiv k\alpha\;\;(\mod\,1)\,,
\equation(dkalpha)
$$
for all energies $E$ in that gap.
The gaps are believed to be open for every $\alpha$.
So far this has been proved for Liouville values [\rCEY,\rAvJit]
and Diophantine values [\rPuig] if $\lambda\ll\lambda'$
or $\lambda'\ll\lambda$.

The spectrum for irrational $\alpha$ is a Cantor set [\rAvJit]
of measure $4|\lambda'-\lambda|$,
as was conjectured in [\rAuAn] and proved later in [\rLasti,\rLastii].
The generalized eigenfunctions
for $\lambda>\lambda'$ are localized in the $n$-direction
and extended in the $m$-direction;
the same holds for $\lambda'>\lambda$, but with $m$ and $n$ exchanged.

The dual Hamiltonian, obtained by interchanging $\lambda$ and $\lambda'$,
is unitarily equivalent to $H^\alpha$.
Of particular interest is the self-dual case $\lambda=\lambda'$.
The spectrum for this family of operators $H^\alpha$,
plotted as points $(\alpha,E)$ in the plane,
is known as the Hofstadter butterfly [\rHof].
The positive-energy part is shown in Figure 1 for $\lambda=\lambda'=1$.
To be more precise, the solid regions are the spectral gaps,
and the color encodes the gap index $k$.
The largest regions are for $k=0$ (white) and $k=\pm 1$.

%%%%%%%%%%%%%%%%%%%%%%%%%%%%%%%%%%%%%%%%%%%%%%%%%%%%%%%%%%%%%
\vskip 0.5cm
\hbox{\hskip0.4cm
\includegraphics[height=5.0cm,width=13cm]{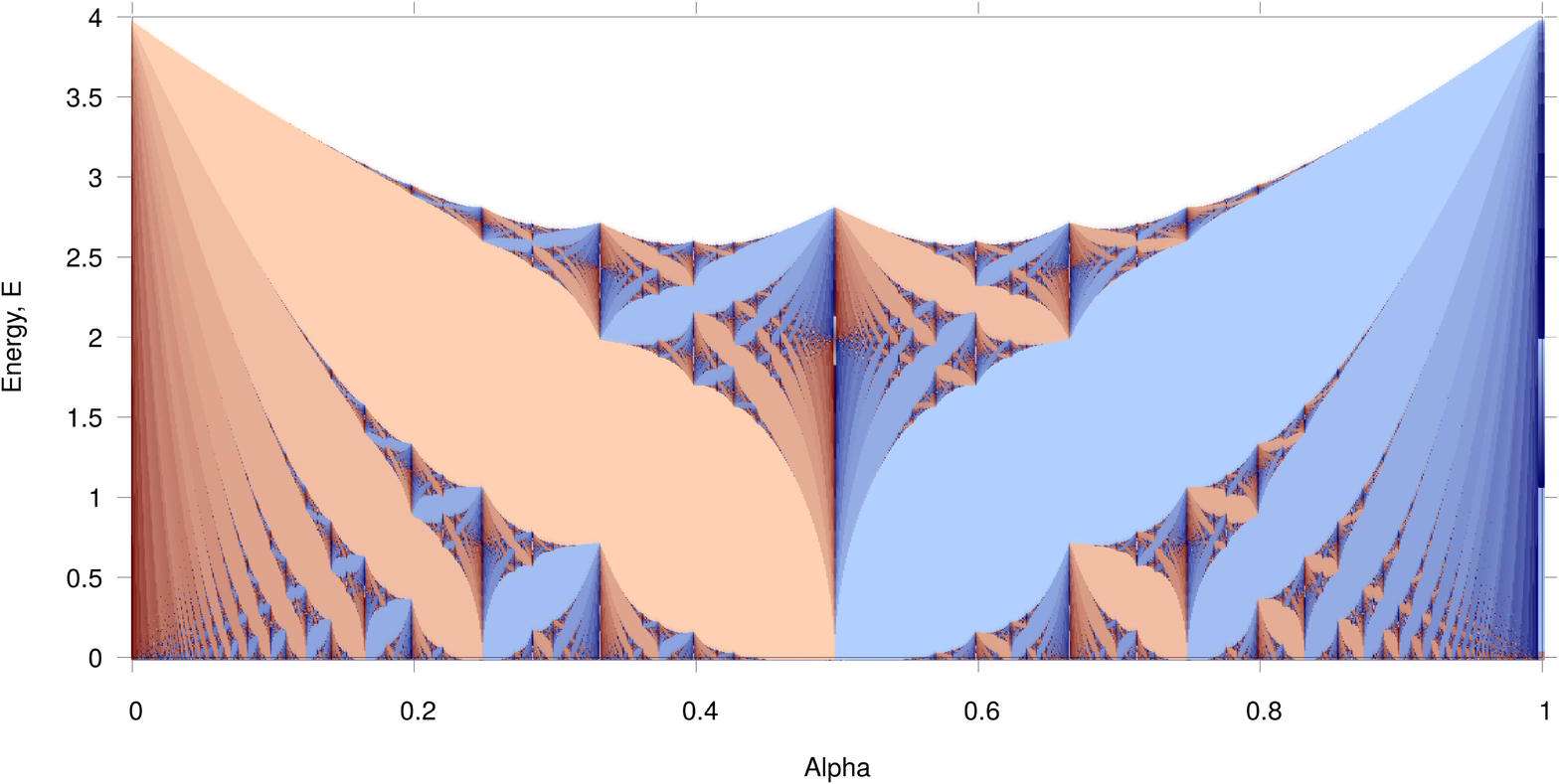}
}
\centerline{\hskip0.0cm\eightpoint{\bf Figure 1.}
Positive-energy part of the Hofstadter butterfly.}
\vskip 0.4cm
%%%%%%%%%%%%%%%%%%%%%%%%%%%%%%%%%%%%%%%%%%%%%%%%%%%%%%%%%%%%%

There exists an extensive literature on the topological
features of the Hofstadter butterfly and their connection with
continued fractions; see e.g.~[\rRuPi,\rATW,\rOsAv,\rSatij] and references therein.
In this paper, we are interested mostly in metric properties,
especially accumulation phenomena in the energy direction,
or for variations in $\lambda$.
Other scaling properties will be described as well.

To simplify the analysis,
we focus mainly on the inverse golden mean $\alpha_\ast=\shalf(\sqrt{5}-1)$.
Figure 2 shows successive magnifications of the Hofstadter butterfly
near the point $(\alpha_\ast,0)$.
As will be described below,
each magnification step is a composition of three basic steps,
and self-similarity occurs with a period $\ell=6$.
Based on high-accuracy computations,
we conjecture that the six-step scaling factor $\mu_1$ is the largest root
of the polynomial $\PP_6$ given below.
Its numerical value is $\mu_1=196.29\ldots$

%%%%%%%%%%%%%%%%%%%%%%%%%%%%%%%%%%%%%%%%%%%%%%%%%%%%%%%%%%%%%
\vskip0.5cm
\hbox{\hskip0.1cm
\includegraphics[height=4.0cm,width=6.8cm]{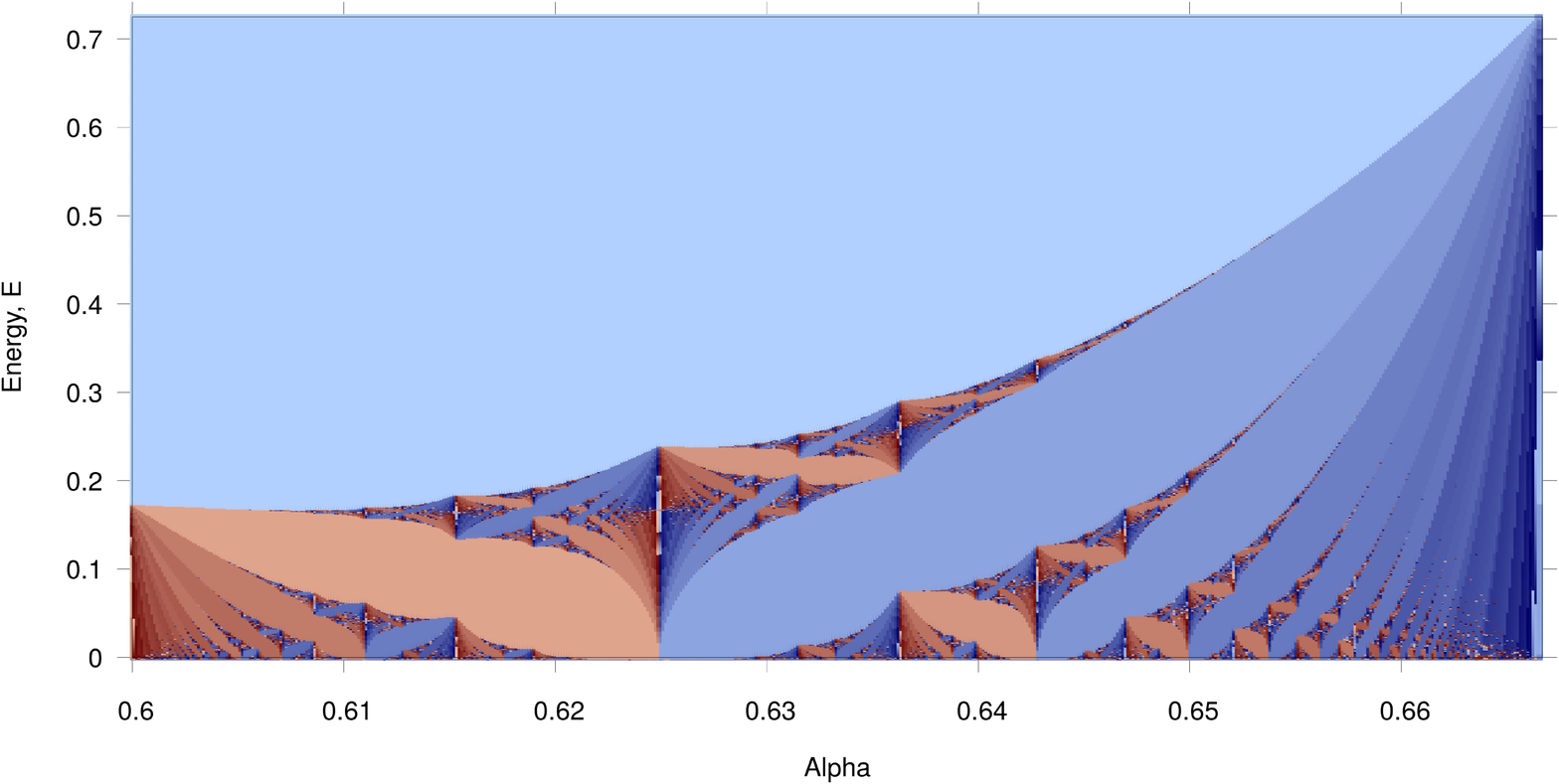}
\includegraphics[height=4.0cm,width=6.8cm]{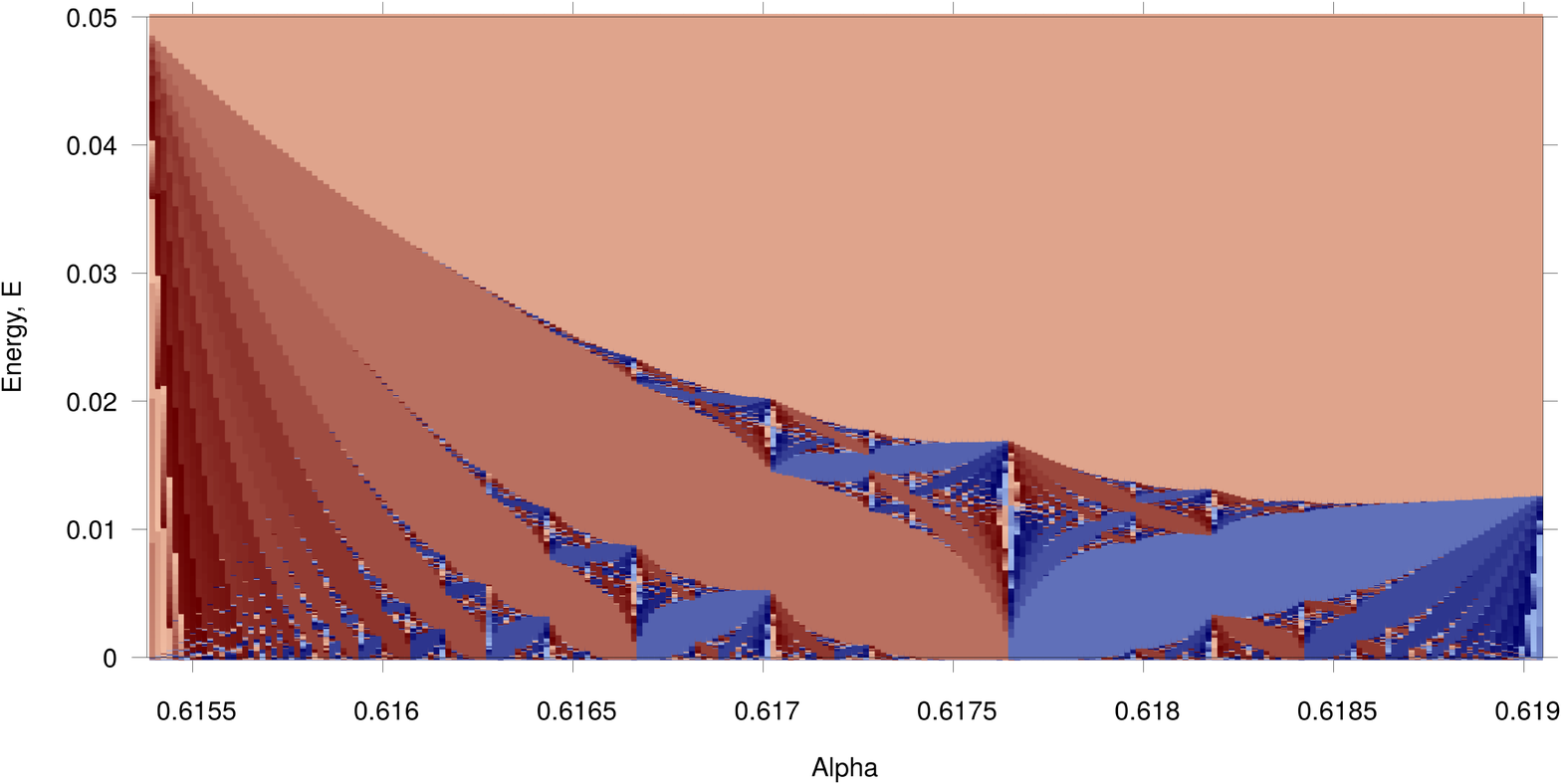}
}
\centerline{\hskip0.0cm\eightpoint{\bf Figure 2.}
$3$-step and $6$-step
enlargements of the Hofstadter butterfly near $(\alpha_\ast,0)$.}
\vskip0.4cm
%%%%%%%%%%%%%%%%%%%%%%%%%%%%%%%%%%%%%%%%%%%%%%%%%%%%%%%%%%%%%

Another scaling of the Hofstadter butterfly was described
in [\rKochAM] near the point $(\alpha_\ast,E_3)$,
where $E_3=2.597\ldots$ is the largest value in the spectrum of $H^{\alpha_\ast}$.
In that case, self-similarity occurs with a period $\ell=3$,
and the three-step scaling factor is $\mu_1=30.790\ldots$
We conjecture that $\mu_1$ is the largest root of $\PP_3$, where
$$
\eqalign{
\PP_3(z)&=z^4-30z^3-24z^2-10z-1\,,\cr
\PP_6(z)&=z^4-196z^3-58z^2-4z+1\,.\cr}
\equation(PiiiPvi)
$$
We note that the product of the two real roots of
$\PP_\ell$ is $(-\alpha_\ast)^{-\ell}$.
Their average $\zeta$
satisfies $\zeta^2-15\zeta-5=0$ in the case $\ell=3$,
and $\zeta^2-98\zeta-19=0$ in the case $\ell=6$.
Using these relations, it is easy to write down explicit expressions
for the real roots of $\PP_\ell$.
Other scaling points $(\alpha_\ast,E_\ell)$, where we expect similar results,
will be described below.
But our computations in those cases are not accurate enough
to yield a good guess for $\PP_\ell$.

\section Skew-product maps
%%%%%%%%%%%%%%%%%%%%%%%%%%

The Hamiltonian $H^\alpha$ commutes with
the dual magnetic translations, one of which is given by
$(\UU_\alpha\phi)(n,m)=\phi(n,m-1)$.
So its spectrum can be determined
by restricting $H^\alpha$ to generalized eigenfunctions
$\phi_\xi(n,m)=e^{-2\pi i m\xi}u_n$ of the translation $\UU_\alpha$.
The restricted Hamiltonian $\HH^\alpha$,
also known as the almost Mathieu ({\ninerm AM}) Hamiltonian,
is defined by $(H^\alpha\phi_\xi)(n,m)=e^{-2\pi im\xi}(\HH^\alpha u)_n$
and takes the form of a Schr\"odinger operator
$$
(\HH^\alpha u)_n=u_{n-1}+u_{n+1}+V(n\alpha)u_n\,,\qquad n\in\integer\,,
\equation(HofToSchroed)
$$
with (quasi)periodic potential $V(x)=2\lambda\cos(2\pi(x+\xi))$.
The equation $\HH^\alpha u=Eu$
for a generalized eigenvector of $\HH^\alpha$ can be written as
$$
\stwovec{u_{n+1}}{u_n}
=A(n\alpha)\stwovec{u_n}{u_{n-1}}\,,\qquad
A(x)=\stwomat{E-V(x)}{-1}{1}{0}\in\,\rmSL(2,\real)\,.
\equation(HHEigenRecursion)
$$
When combined with a rotation $x\mapsto x+\alpha$
of the circle $\torus=\real/\integer$,
this recursion defines a skew-product map $G$,
$$
G(x,y)=(x+\alpha,A(x)y)\,,\qquad
x\in\torus\,,\quad y\in\real^2\,.
\equation(AMG)
$$
Two dynamical quantities
of interest here are the Lyapunov exponent $L(G)$
and the fibered rotation number $\varrho(G)$. They can be defined as follows.
Let $\GG$ be a lift of the map
$(x,y)\mapsto\bigl(x+\alpha,\|A(x)y\|^{-1}A(x)y\bigr)$
from $\torus\times\circle$ to $\torus\times\real$,
where $\circle$ denotes the unit circle $\|y\|=1$ in $\real^2$.
Then
$$
L(G)=\lim_{n\to\infty}{1\over n}\log\bigl\|(\mat G^n)(x)\bigr\|\,,\qquad
\varrho(G)=\lim_{n\to\infty}{1\over 2\pi n}\arg\GG^n(x,\vartheta)\,.
\equation(LyapRot)
$$
Here, $\mat G^n$ denotes the matrix part of $G^n$,
and $\arg(x,\vartheta)=\vartheta$.
Assuming that $A:\torus\to\rmSL(2,\real)$ is continuous and $\alpha$ irrational,
the limit for $\varrho(G)$ exists,
is independent of $x$ and $\vartheta$,
independent modulo $1$ of the choice of the lift $\GG$,
and convergence is uniform.
Under the same assumptions, the limit for $L(G)$
exists and is a.e.~constant in $x$.
If $G$ is an \AM skew-product for energy $E$,
then the fibered rotation number is related
to the density of states via $d(\alpha,E)\equiv-2\varrho(G)$ modulo $1$.
Furthermore, $L(G)=\max(0,\log|\lambda|)$,
if $E$ belongs to the spectrum of $\HH^\alpha$.
For proofs of these facts we refer to [\rBeSi,\rJoMo,\rAvSi,\rGoSch].

Figure 3 depicts two scaling properties of the \sAM map $G$ for energy $E$
near the above-mentioned points $E_\ell$ in the spectrum of $H^{\alpha_\ast}$.
The graph on the left shows the logarithm of the Lyapunov exponent
$f(E)=L(G)$ as a function of the logarithm of
$\epsilon=|E-E_\ell|$, in the case $\ell=3$.
Based on our renormalization analysis described below,
we expect that
$$
f(E)\simeq C_{\pm}(\log\epsilon)\epsilon^\tau\,,\qquad
\tau={\ell\log(\alpha_\ast^{-1})\over\log(\mu_1)}\,,
\equation(LyapFiberedScaling)
$$
as $E\to E_\ell$ from below ($-$) or from above ($+$).
The functions $C_{\pm}$ are periodic with period $\log(\mu_1)$,
where $\mu_1$ is the largest root of $\PP_\ell$.
The same behavior is observed for $\ell=6$.
In the case $\ell=3$, the function $C_{+}$ is constant,
since $H^{\alpha_\ast}$ has no spectrum above $E_3$.

The graph in the right part of Figure 3 shows the logarithm of
$f(E)=2|\rho(E)-\rho(E_\ell)|$ versus $\log\epsilon$
in the case $\ell=6$.
Here, $\rho(E)$ denotes the fibered rotation number
of the \sAM map with energy $E$.
The predicted behavior is again of the form \equ(LyapFiberedScaling),
both for $\ell=3$ and $\ell=6$.
The value of $\tau$ given by \equ(LyapFiberedScaling),
as well as the period of $C_{\pm}$,
match our numerical data at the precision available.

%%%%%%%%%%%%%%%%%%%%%%%%%%%%%%%%%%%%%%%%%%%%%%%%%%%%%%%%%%%%%
\vskip0.5cm
\hbox{\hskip0.2cm
\includegraphics[height=4.0cm,width=6.0cm]{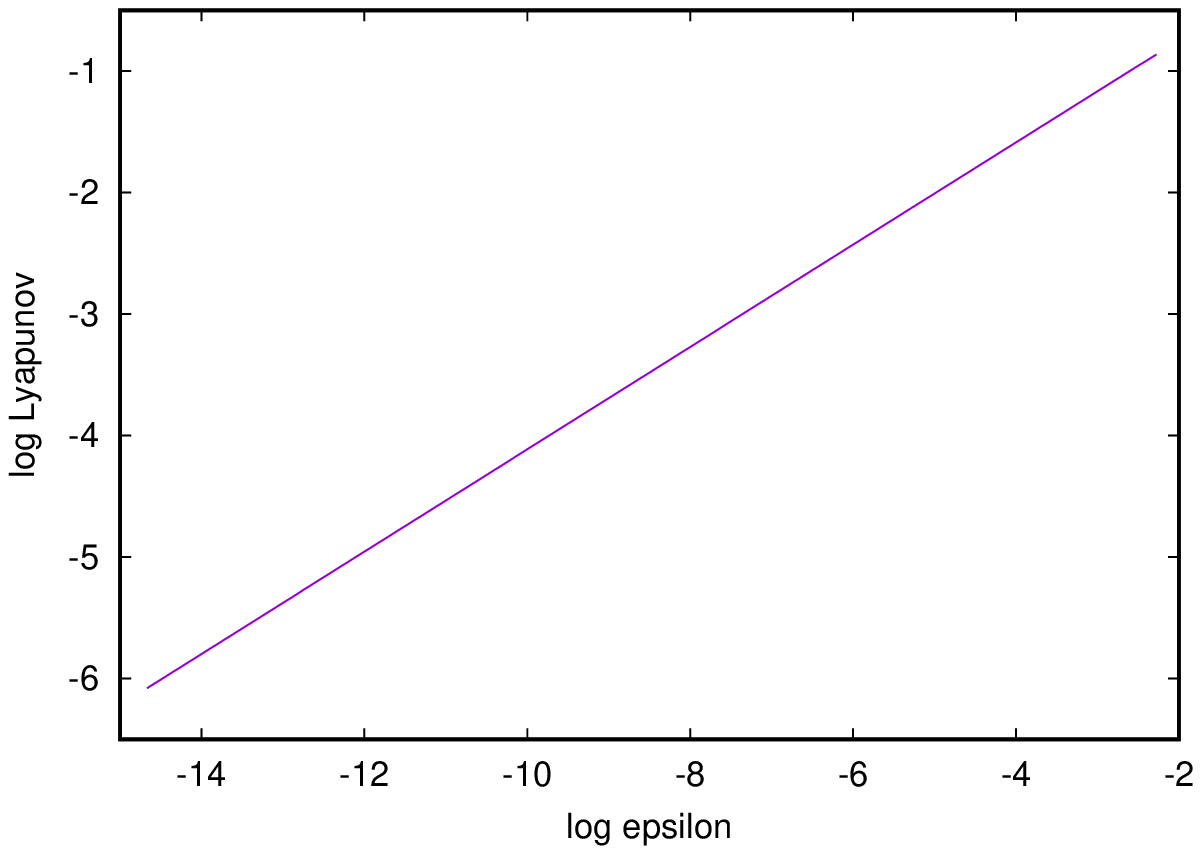}
\includegraphics[height=4.0cm,width=7.5cm]{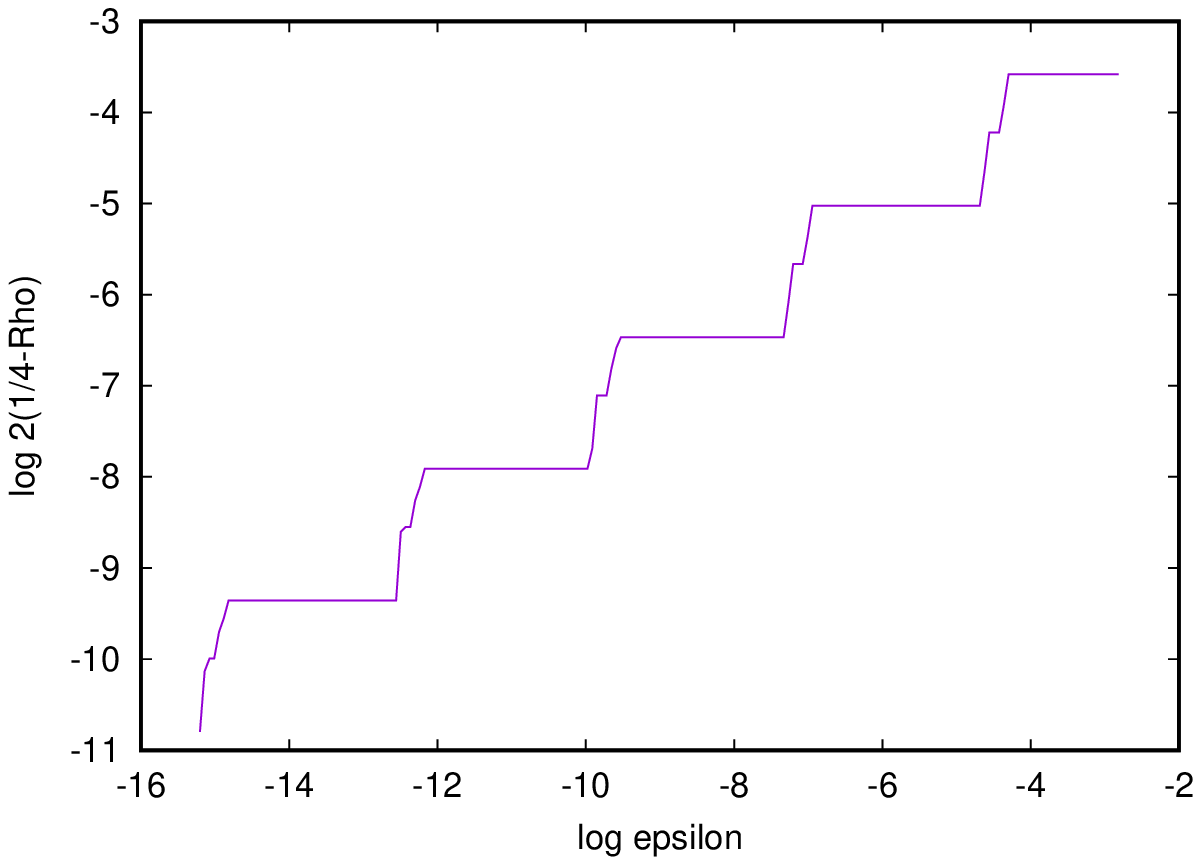}
}
\centerline{\hskip0.0cm\eightpoint{\bf Figure 3.}
Scaling of $L$ for $E>E_3$ near $E_3$ (left)
and of $1/4-\varrho$ for small $E>E_6=0$ (right).}
\vskip0.0cm
%%%%%%%%%%%%%%%%%%%%%%%%%%%%%%%%%%%%%%%%%%%%%%%%%%%%%%%%%%%%%

\section Renormalization
%%%%%%%%%%%%%%%%%%%%%%%%

The observation of such asymptotic scaling suggests
that a suitable renormalization transformation $\buR$
for skew-product maps has a periodic point $P_\ast$
of period $\ell$, and that $\mu_1$ is the largest
eigenvalue of the derivative $D\buR^\ell(P_\ast)$.
Let $G$ be a skew-product map as in \equ(AMG),
henceforth abbreviated as $G=(\alpha,A)$.
Regard $G$ as a map on $\real\times\real^2$.
If $G$ arises from a Schr\"odinger operator \equ(HofToSchroed)
with a $1$-periodic potential $V$,
then we pair $G$ with a second skew-product map $F=(1,\id)$.
The $1$-periodicity of the matrix function $A$ is expressed by the fact that $G$
commutes with $F$.
As was observed and used in [\rKochAM], the \AM map $G$ with potential
$V(x)=2\lambda\cos(2\pi(x+\xi))$ and $\xi=\alpha/2$
is reversible, in the sense that
$$
G^{-1}=\SS G\SS\,,\qquad
\SS(x,y)=(-x,Sy)\,,\qquad S=\stwomat{0}{1}{1}{0}\,.
\equation(Reversible)
$$
Thus, we restrict now to pairs $P=(F,G)$ that are reversible,
and we choose our renormalization transformation
to preserve reversibility, provided that $F$ and $G$ commute.
The matrix parts of $F$ and $G$ are always
assumed to take values in $\rmSL(2,\real)$.
The transformation $\buR$ considered in [\rKochAM] is given by
$$
\buR(P)=\bigl(\Lambda_1^{-1}G\Lambda_1\,\bcomma\,
\Lambda_1^{-1}FG^{-1}\Lambda_1\bigr)\,,\qquad
\Lambda_1(x,y)=\bigl(\alpha_\ast x,Se^{\sigma_1 S}y\bigr)\,,
\equation(RGOne)
$$
where $\sigma_1=\sigma_1(P)$ is determined by a suitable
normalization condition.
We note that this choice of $\buR$ is tailored
to the study of skew-product maps $G=(\alpha,A)$ with $\alpha=\alpha_\ast$.
Due to the identity $1-\alpha_\ast=\alpha_\ast^2$,
a pair $((1,B),(\alpha_\ast,A))$ is mapped to a pair $((1,B_1),(\alpha_\ast,A_1))$.
Analogous transformations can be defined for other quadratic
irrationals $\alpha$.
Approximate renormalization schemes
and limiting cases have been considered earlier in [\rMeOsWi,\rDaMe].

It is instructive to consider what happens to fibered
rotation numbers under renormalization.
Let $F=(\beta,B)$ and $G=(\alpha,A)$,
with $\alpha$ and $\beta$ irrational.
Assume that $A$ and $B$ are continuous $1$-periodic functions on $\real$,
taking values in $\rmSL(2,\real)$.
If $F$ and $G$ commute, then $\varrho(FG)\equiv\varrho(F)+\varrho(G)$ modulo $1$.
This follows e.g.~from the uniform convergence [\rJoMo]
of the second limit in \equ(LyapRot).
As a consequence, if $(F_1,G_1)=\buR(F,G)$, then
$$
\twovec{\varrho(F_1)}{\varrho(G_1)}
\equiv\twomat{0}{1}{1}{-1}
\twovec{\varrho(F)}{\varrho(G)}\quad(\mod\,1)\,.
\equation(RotRG)
$$
This equation defines a hyperbolic map on the torus $\torus^2$,
related to Arnold's cat map [\rDyFa].
It has a dense set of periodic orbits,
with homoclinic or heteroclinic connections between any two them.
In particular,
every point $\bigl[{0\atop m/n}\bigr]$ lies on a periodic orbit.
If $m$ and $n$ are coprime,
then its period agrees with the fundamental period $\ell(n)$
of the Fibonacci sequence modulo $n$.
To see why, multiply both rotation vectors in \equ(RotRG) by $n$,
to get a congruence modulo $n$ over the integers.
Denoting by $U$ the $2\times 2$ matrix in \equ(RotRG),
the condition for a period $\ell$ is $U^\ell\equiv\id$ modulo $n$.
A straightforward computation shows that this condition holds
if and only if $\ell$ is a period of the Fibonacci sequence modulo $n$.
The smallest such integer $\ell>0$ is known
as the Pisano period $\ell(n)$.
The two periods described earlier are $\ell(2)=3$ and $\ell(4)=6$.
Periods $\ell(n)$ with odd $n$ are not expected
to occur in the \AM model, due to a symmetry that implies
$\varrho(E)+\varrho(-E)\equiv\shalf$ modulo $1$.

For even $n$, the period $\ell(n)$ is a multiple of $3$.
Thus, we restrict our analysis to iterates $\buR^\ell$ with $\ell$ a multiple of $3$.
Notice that $\buR^\ell(P)$ can be obtained by
first iterating $(F,G)\mapsto(G,FG^{-1})$ $\ell$ times,
and then conjugating the resulting maps with a scaling
$$
\Lambda_\ell(x,y)=\bigl(\alpha_\ast^\ell x,S^\ell e^{\sigma_\ell S}y\bigr)\,,
\equation(Lambdaell)
$$
where $\sigma_\ell=\sigma_\ell(P)$ is determined
by a suitable normalization condition.

The following result is a slight extension of Theorem 1.1 in [\rKochAM].

\claim Theorem(ExistenceThree)
$\buR^3$ has a reversible fixed point $P_\ast=(F_\ast,G_\ast)$
with $F_\ast$ and $G_\ast$ commuting.
$P_\ast$ is not a fixed point of $\buR$.
The matrix parts of $F_\ast$ and $G_\ast$ are non-constant
and extend to entire analytic functions.
The scaling exponent $\sigma_\ast=\sigma_3(P_\ast)$ is positive
and satisfies the bound $|\sigma_\ast-c_3|<10^{-443}$,
where $c_3=\shalf\cosh^{-1}(\alpha_\ast^{-1})$.

%% used PDeg =  767, Precision =  1536
%% see check_fixpt.log7

We conjecture that $\sigma_\ast=c_3$ and note that
the squared $y$-scaling factors $e^{\pm 2c_3}$
are the real roots of the polynomial $\QQ_3(z)=z^4-2z^3-2z^2-2z+1$.

The following theorem is proved in [\rKK].

\claim Theorem(ExistenceSix)
$\buR^6$ has a reversible fixed point $P_\ast=(F_\ast,G_\ast)$
with $F_\ast$ and $G_\ast$ commuting.
$P_\ast$ is not a fixed point of $\buR^k$ for any positive $k<6$.
The matrix parts of $F_\ast$ and $G_\ast$ are non-constant
and extend to entire analytic functions.
The scaling exponent $\sigma_\ast=\sigma_6(P_\ast)$ is positive
and satisfies the bound $|\sigma_\ast-c_6|<10^{-431}$,
where $c_6=\shalf\cosh^{-1}(\alpha_\ast^{-3})$.

%% used PDeg =  767, Precision =  1536
%% see check_fixpt6.log4

We conjecture that $\sigma_\ast=c_6$
and note that $e^{\pm 2c_6}$ are the real roots of
the polynomial $\QQ_6(z)=z^4-8z^3-2z^2-8z+1$.

The fixed point $P_\ast$ described in
\clm(ExistenceThree) (\clm(ExistenceSix))
can be associated with the Pisano period $\ell=\ell(n)$
for $n=2$ ($n=4$).
Numerically, $P_\ast$ attracts the \sAM pair with energy $E_\ell$
under iteration of $\buR^\ell$.
In addition, we have numerical evidence for
the existence of analogous fixed points for $n=6$ and $n=8$.
The corresponding rotation number is $\varrho(G)=\sfrac{1}{n}$
in all cases considered.
The periods for $n=6,8$ are $\ell(n)=24,12$,
and the corresponding energy values are
$E_{12}=1.990\ldots$ and $E_{24}=1.888\ldots$, respectively.
Our computations for $n=8$ were carried out
at sufficient accuracy to predict a scaling exponent
$c_{12}=\shalf\cosh^{-1}(\alpha^{-6})$.
This was motivated by the observation that $c_6=2c_3$.
A similar relation for $c_{12}$ seems excluded.

\demo Remark(OtherPeriod)
For convenience we have labeled periodic orbits
by their fundamental period $\ell$.
However, the torus map \equ(RotRG) can have several periodic orbits
with fundamental period $\ell(n)$.
They arise from Fibonacci integer sequences modulo $n$
that do not include a consecutive pair $(0,1)$.
Furthermore, there could be more than one (or no) periodic orbit of $\buR$
for some periodic orbits of the map \equ(RotRG).

\clm(ExistenceThree) is proved by first solving
the fixed point problem for the transformation
$$
\buR_3(F,G)=(\Lambda_3^{-1}GF^{-1}G\Lambda_3\,\bcomma\,
\Lambda_3^{-1}G^{-1}FG^{-1}FG^{-1}\Lambda_3)\,,
\equation(PaliRGThree)
$$
which is obtained from $\buR^3$
by a ``palindromic'' re-arrangement of the factors $F^{\pm 1}$ and $G^{\pm 1}$.
This transformation has the advantage that reversible pairs
are mapped to reversible pairs, even if the component maps do not commute.
After establishing the existence of a fixed point $P_\ast$ for $\buR_3$,
we prove that its components $F_\ast$ and $G_\ast$ commute.
An analogous approach is used in our proof of \clm(ExistenceSix).

We note that, due to the scaling $x\mapsto\alpha_\ast x$ involved,
the analysis can be carried out on a bounded domain in $\complex$.
Entire analyticity of the matrix functions $B_\ast=\mat F_\ast$
and $A_\ast=\mat G_\ast$ follows by iterating the fixed point equation
and using that $x\mapsto\alpha_\ast x$ is analyticity improving.
The same argument shows that $A_\ast$ and $B_\ast$
are exponentially bounded on all of $\complex$.
More specific bounds can be obtained by using
information on the Lyapunov exponent of maps that
are attracted to $P_\ast$ under renormalization.
Based on an explicit expression [\rAvG]
for the Lyapunov exponent of complex-translated \AM maps,
we expect that $\log\|B_\ast(x)\|$ and $\log\|A_\ast(\alpha_\ast x)\|$
grow like $2\pi 5^{-1/2}|x|$ in the imaginary direction.
This is consistent with the decay rate of the Taylor coefficient
that we find numerically in the cases $\ell=3$ and $\ell=6$.

\section Scaling and universality
%%%%%%%%%%%%%%%%%%%%%%%%%%%%%%%%%

In both cases ($\ell=3$ and $\ell=6$) our analysis
requires as input an approximate fixed point of $\buR^\ell$.
Such a pair $P_{k\ell}$ is obtained numerically
by starting with the \sAM pair $P$ with energy $E_\ell$
and computing $P_{k\ell}=\buR^{k\ell}(P)$ for some large $k$.
The fact that this procedure works suggests that $P$ is attracted
to our fixed point $P_\ast$ under the iteration of $\buR^\ell$.
If we assume that this is the case,
then it is possible to relate asymptotic properties
of $P$ to local properties of the transformation $\buR$
near the orbit of $P_\ast$.
Since $\buR$ defines a dynamical system on a space of pairs,
the same applies to other families in the domain of $\buR$.

\smallskip
Consider e.g.~a Schr\"odinger operator $\HH^{\alpha_\ast}$
and the associated map $G=(\alpha_\ast,A)$.
Let $P=((1,\id),G)$ and $(F_n,G_n)=\buR^n(P)$.
Then $G_n=\Lambda_n^{-1}G^{q_n}\Lambda_n$,
where $q_n$ is the $n+1^{\rm st}$ Fibonacci number.
The matrix part of $G^{q_n}$
is related to the matrix part $A_n$ of $G_n$ via
$$
\bigl(\mat G^{q_n}\bigr)(\alpha_\ast^n x)
=e^{\sigma_n S}S^n A_n(x)S^ne^{-\sigma_n S}\,.
\equation(AAqnAn)
$$
If the sequence $k\mapsto P_{k\ell}$
converges to a fixed point $P_\ast$ of $\buR^\ell$,
then $\sigma_{k\ell}\sim k\sigma_\ast$ for large $k$,
where $\sigma_\ast$ is the scaling exponent associated with $P_\ast$.
This shows that the scaling factors $e^{\sigma_\ast}$
given in Theorems \clmno(ExistenceThree) and \clmno(ExistenceSix)
describe the asymptotic behavior
of generalized eigenfunctions of $\HH^{\alpha_\ast}$
with the proper rotation numbers.
A precise argument along these lines is given in [\rKochAM],
as well as a graph of the generalized eigenfunction for the \sAM
Hamiltonian for energy $E_3$.

Concerning proper rotation numbers, we note that,
while the periodic orbits of the map \equ(RotRG) are pairs with rational components,
their stable manifolds include mostly irrational pairs.
In particular, all pairs with $\varrho(F)=0$
and $2\varrho(G)\in\integer[\alpha_\ast]$ are attracted to rational periodic orbits.
Numerically, we find e.g.~that the \sAM pair
with $2\varrho(G)=1-\alpha_\ast$ is attracted under iteration of $\buR$
to the $3$-periodic orbit described in \clm(ExistenceThree).
The corresponding energy is $E=1.874\ldots$.

\smallskip
Other universal quantities are associated
with the eigenvalues of modulus $\ge 1$
of the derivative of $\buR^\ell$ at a fixed point $P_\ast$.
These eigenvalues have been determined numerically
for the fixed points described in
Theorems  \clmno(ExistenceThree) and \clmno(ExistenceSix).
In both cases, $\buR^\ell$ appears to be hyperbolic,
with exactly two eigenvalues of modulus $\ge 1$.
It seems likely that the same is true much more generally.
In some sense, the ``universality class''
is governed by the two-parameter
\hskip-2pt\footnote{$^1$}
{\eightpoint\hskip-2.7em
We are not counting here the parameter $\alpha$;
any scaling in the $\alpha$ direction is trivial
and determined solely by arithmetic.}
\AM model.
To be more precise about hyperbolicity:
We restrict $\buR^\ell$ to a codimension $1$ manifold
that includes all commuting pairs in the space being considered.
Without this restriction,
$D\buR^\ell(P_\ast)$ has a simple eigenvalue $(-1)^\ell$
that is associated with a non-commuting perturbation of $P_\ast$.

Two scaling phenomena that are governed by the largest eigenvalue $\mu_1$
are described by \equ(LyapFiberedScaling).
This eigenvalue determines the $\ell$-step asymptotic scaling
of the Hofstadter butterfly at $(\alpha_\ast,E_\ell)$ in the energy direction.
One of the assumptions here is that the \AM family
intersects the stable manifold of $\buR^\ell$ transversally;
or equivalently, that this family
converges to the unstable manifold of $P_\ast$
under the iteration of $\buR^\ell$ and proper rescaling.
To be more precise,
let $\mu_2$ be the second largest eigenvalue of $D\buR^\ell(P_\ast)$,
and define $\mu s=(\mu_1s_1,\mu_2s_s)$ for all $s=(s_1,s_2)$ in $\real^2$.
For $s$ near zero,
denote by $P(s)$ the \AM pair for $E=E_\ell+s_1$ and $\lambda=1+s_2$.
Then the family $s\mapsto\buR^{k\ell}(P(\mu^{-k}s))$
is assumed to converge to a parametrization of the local
unstable manifold of $\buR^\ell$ at $P_\ast$ as $k$ tends to infinity.
We note that the diagonal nature of the
parameter-scaling $\mu$ is specific to the \AM family.

Consider e.g.~the one-parameter family obtained by setting $s_2=0$.
Assuming that the Lyapunov exponents and rotation numbers
have limits as well, a straightforward computation
yields the behavior \equ(LyapFiberedScaling) for positive $\epsilon=|s_1|$
close to zero.

Next, consider the one-parameter family obtained by setting $s_1=0$.
Let $s_2>0$, and denote by $G(s_2)$ the second component of the pair $P(s)$.
In this case, we already know that $L(G(s_2))=\log(1+s_2)$.
So the assumptions made above yield a prediction
for the eigenvalue $\mu_2$.
Notice that, up to a conjugacy by $\Lambda_{k\ell}$,
the second component of $\buR^{k\ell}(P(\mu^{-k}s))$
is the map $G\bigl(\mu_2^{-k}s_2\bigr)^{q_{k\ell}}$,
where $q_n\simeq 5^{-1/2}\alpha_\ast^{-n-1}$
denotes the $n+1^{\rm st}$ Fibonacci number.
Assuming that the Lyapunov exponent of $G\bigl(\mu_2^{-k}s_2\bigr)^{q_{k\ell}}$
converges to a finite nonzero value as $k\to\infty$,
the above implies that $\mu_2=\alpha_\ast^{-\ell}$.
This value of $\mu_2$ is indeed observed numerically,
for the two periods $\ell(2)=3$ and $\ell(4)=6$.

We have also computed the first $12$ contracting eigenvalues
of $\LL=D\buR_3^{\ell/3}(P_\ast)$.
Numerically, the fifth largest (in modulus) eigenvalue $\mu_5$
is a real root of the polynomial $\PP_\ell$,
related to largest eigenvalue $\mu_1$ as described after \equ(PiiiPvi).
For these real roots $\mu$ of $\PP_\ell$,
one also finds that $x=\mu^{1/3}$
satisfies $x^4-3x^3-x-1=0$ in the case $\ell=3$,
while $x=\mu^{1/2}$
satisfies $x^4-14x^3-2x-1=0$ in the case $\ell=6$.
The remaining eigenvalues of $\LL$ appear to be (real and) of the form
$(\pm\alpha_\ast)^{k\ell}$, $(\pm\alpha_\ast)^{k\ell}e^{2c_\ell}$,
or $(\pm\alpha_\ast)^{k\ell}e^{-2c_\ell}$, for some positive integer $k$.
Both signs appear, if we count multiplicities in the case $\ell=6$.
For at least one choice of the sign,
the eigenvector is generated by a coordinate transformation
or corresponds to a non-commuting direction;
see [\rKochAM] for details on how to determine these eigenvectors.
For the other values we do not have an explanation.

\bigskip\noindent
{\bf Acknowledgments}.
The work of S.K.~is supported in part
by the National Science Foundation EPSCoR RII Track-4 Award No.~1738834.

%%%%%%%%%%%%%%%%%%%%%%%%%%%%%%%%%%%%%%%%%%%%%%%%%%
\bigskip
\references
%%%%%%%%%%%%%%%%%%%%%%%%%%%%%%%%%%%%%%%%%%%%%%%%%%

{\ninepoint

\item{[\rHarp]} P.G.~Harper,
{\sl Single band motion of conduction electrons in a uniform magnetic field},
Proc. Phys. Soc. Lond. A {\bf 68}, 874--892 (1955).

\item{[\rHof]} D.R.~Hofstadter,
{\sl Energy levels and wave functions of Bloch electrons
in rational and irrational magnetic fields},
Phys. Rev. B 14, 2239--2249 (1976).

\item{[\rAuAn]} S.~Aubry, G.~Andr\'e,
{\sl Analyticity breaking and Anderson localization in incommensurate lattices},
Ann. Israel Phys. Soc. {\bf 3}, 133--164 (1980).

\item{[\rBeSi]} J.~Bellissard, B.~Simon,
{\sl Cantor spectrum for the almost Mathieu equation},
J. Funct. Anal. {\bf 48}, 408--419 (1982).
%\pdfclink{0 0 1}{online here}
%{http://citeseerx.ist.psu.edu/viewdoc/download?doi=10.1.1.211.2874&rep=rep1&type=pdf}

\item{[\rTKNdN]} D.J.~Thouless, M.~Kohmoto, M.P.~Nightingale, M.~den Nijs,
{\sl Quantized Hall conductance in a two-dimensional periodic potential},
Phys. Rev. Lett. {\bf 49}, 405--408 (1982).
%\pdfclink{0 0 1}{online here}
%{https://journals.aps.org/prl/pdf/10.1103/PhysRevLett.49.405}

\item{[\rJoMo]} R.~Johnson, J.~Moser,
{\sl The rotation number for almost periodic potentials},
Commun. Math. Phys. {\bf 84}, 403--438 (1982).
%\pdfclink{0 0 1}{online here}
%{https://projecteuclid.org/euclid.cmp/1103921211}

\item{[\rAvSi]} J.~Avron, B.~Simon,
{\sl Almost periodic Schr\"odinger operators. II.
The integrated density of states},
Duke Math. J. {\bf 50}, 369--391 (1983).
%\pdfclink{0 0 1}{online here}
%{https://www.math.caltech.edu/SimonPapers/149.pdf}

\item{[\rDyFa]} F.J.~Dyson, H.~Falk,
{\sl Period of a discrete cat mapping},
Amer. Math. Monthly {\bf 99}, 603--614 (1992).

\item{[\rLasti]} Y.~Last,
{\sl A relation between a.c.~spectrum of ergodic Jacobi matrices
and the spectra of periodic approximants},
Commun. Math. Phys. {\bf 151}, 183--192 (1993).
%\pdfclink{0 0 1}{online here}
%{https://projecteuclid.org/euclid.cmp/1104252049}

\item{[\rLastii]} Y.~Last,
{\sl Zero measure spectrum for the almost Mathieu operator},
Comm. Math. Phys. {\bf 164}, 421--432 (1994).
%\pdfclink{0 0 1}{online here}
%{https://projecteuclid.org/download/pdf_1/euclid.cmp/1104270838}

\item{[\rRuPi]} A.~R\"udinger, F.~Pi\'echon,
{\sl Hofstadter rules and generalized dimensions of the spectrum of Harper's equation},
J. Phys. A {\bf 30}, 117--128 (1997).
%\pdfclink{0 0 1}{online here}{https://arxiv.org/pdf/cond-mat/9610177}

\item{[\rATW]} A.G.~Abanov, J.C.~Talstra, P.B.~Wiegmann,
{\sl Hierarchical structure of Azbel-Hofstadter problem:
strings and loose ends of Bethe ansatz},
Nuclear Physics B {\bf 525}, 571--596 (1998).
%\pdfclink{0 0 1}{online here}
%{https://www.sciencedirect.com/science/article/pii/S0550321398003460/pdf}

\item{[\rMeOsWi]} B.D.~Mestel, A.H.~Osbaldestin, B.~Winn,
{\sl Golden mean renormalisation for the Harper equation:
the strong coupling fixed point},
J. Math. Phys. {\bf 41}, 8304--8330 (2000).
%\pdfclink{0 0 1}{online here}
%{https://dspace.lboro.ac.uk/dspace-jspui/bitstream/2134/747/1/00-07.pdf}

\item{[\rOsAv]} D.~Osadchy, J.E.~Avron,
{\sl Hofstadter butterfly as quantum phase diagram},
J. Math. Phys. {\bf 42}, 5665--5671 (2001).
%\pdfclink{0 0 1}{online here}
%{http://phsites.technion.ac.il/wp-content/uploads/sites/3/2013/05/J_Math_Phys_42_5665_2001_butterfly.pdf}

\item{[\rCEY]} M.D.~Choi, G.A.~Eliott, N.~Yui,
{\sl Gauss polynomials and the rotation algebra},
Invent. Math. {\bf 99}, 225--246 (2002).

\item{[\rDaMe]} J.~Dalton, B.D.~Mestel,
{\sl Renormalization for the Harper equation for quadratic irrationals},
J. Math. Phys. {\bf 44}, 4776--4783 (2003).

\item{[\rPuig]} J.~Puig,
{\sl Cantor spectrum for the almost Mathieu operator},
Commun. Math. Phys. {\bf 244}, 297--309 (2004).
%\pdfclink{0 0 1}{online here}
%{https://mat.upc.edu/en/people/joaquim.puig/articles/puig-cantor-spectrum.pdf}

\item{[\rGoSch]} M. Goldstein and W. Schlag,
{\sl Fine properties of the integrated density of states
and a quantitative separation property of the Dirichlet eigenvalues},
Geom. Funct. Anal. {\bf 18}, 755--869 (2008).

\item{[\rAvJit]} A.~Avila, S.~Jitomirskaya,
{\sl The ten martini problem},
Ann. Math. {\bf 170}, 303--342 (2009).
%\pdfclink{0 0 1}{online here}
%{http://annals.math.princeton.edu/wp-content/uploads/annals-v170-n1-p08-p.pdf}

\item{[\rAvG]} A.~Avila,
{\sl Global theory of one-frequency Schr\"odinger operators},
Acta Math. {\bf 215}, 1--54 (2015).

\item{[\rSatij]} I.I.~Satija,
{\sl A tale of two fractals:
the Hofstadter butterfly and the integral Apollonian gaskets},
Eur. Phys. J. Spec. Top. {\bf 225}, 2533--2547 (2016).
%\pdfclink{0 0 1}{online here}{https://arxiv.org/pdf/1606.09119}

\item{[\rKochAM]} H.~Koch,
{\sl Golden mean renormalization
for the almost Mathieu operator and related skew products},
Preprint 2019, mp\_arc 19-45, arXiv:1907.06804 [math-ph].
%\pdfclink{0 0 1}{online here}
%{https://web.ma.utexas.edu/users/koch/papers/skewrg/}

\item{[\rKK]} H.~Koch, S.~Koci\'c,
{\sl Orbits under renormalization of skew-product maps
over circle rotations},
In preparation.

}

\bye